\documentclass[preprint,showpacs,preprintnumbers,amsmath,amssymb]{revtex4}

\usepackage{graphicx}
\usepackage{dcolumn}
\usepackage{bm}

\begin{document}

\title{Possible Systematic Effects in Fomblin Coated Storage Cell Neutron Lifetime Measurements}

\author{Steve K. Lamoreaux}

\affiliation{Yale University, Department of Physics, P.O. Box 208120, New Haven, CT 06520-8120}

\date{Dec. 4, 2006}

\begin{abstract}
A new neutron lifetime experiment employing frozen Fomblin has produced a result in significant disagreement with previous experiments that used liquid Fomblin near room temperature.  This new experiment is subject to very few corrections, so the source of the discrepancy remains to be identified .  Here we theoretically investigate several possible systematic effects
for near-room-temperature storage experiments. By considering the combined effect of gravity with the usual ultracold neutron losses together with surface wave scattering loss and ultracold spectral evolution, a correction to a previous neutron lifetime experiment is found to be insignificantly small.  The source of the systematic variation between experiments remains to be identified.
\end{abstract}
 \maketitle

\section{Introduction}

A recent determination of the neutron lifetime by Serebrov et al. have yielded a value of $878.5\pm 0.7_{\rm stat}\pm0.3_{\rm sys}$ s \cite{sere}, which is in significant disagreement with the PDG recommended value of $885.7\pm0.8$ s \cite{pdg}.  The purpose of this note is to report the results of a theoretical study of the effects of surface wave scattering on the gravitational corrections to one of the moderate-accuracy experiments \cite{mampe} that is included in the PDG determination.  These results might also have implications for the most accurate experiment because the effects described here have not been included in the modelling of the experimental system \cite{aruz}.

The use of liquid Fomblin, a perfluorinated ether with very low nuclear absorption, to produce a nearly perfect UCN storage bottle was suggested and pioneered by C. Bates \cite{bates}. The highest accuracy storage experiments have employed Fomblin coated storage bottles.
In order to get rid of the thermal scattering, which is a dominant loss near room temperature, \cite{sere} employed frozen Fomblin and achieved a loss per wall reflection of less than $5\times 10^{-6}$, a record. However, this is an order of magnitude larger than expected and is a source of concern.

Neutron lifetime experiments employing the storage of ultracold neutrons (UCN) are generally based measuring the total storage lifetime, which includes both beta decay and wall losses, as a function of volume (or mean free path between collisions) and extrapolating to infinite volume.  In principle, this seems like a good idea; in practice, however, there is a distribution of neutron velocities and each velocity group has a different lifetime.  To get around this problem, in \cite{mampe}, the volume and storage times were adjusted so that the same number of wall collisions occurred for each volume. Because the times were different, the neutron lifetime could in principle be directly determined, independent of the specific loss mechanism.  However, the effects of gravity spoil the simple mean free path scaling because the wall collisions are not identical on all the internal storage cell surfaces; given that the neutron potential is 1.04 neV/cm in the Earth's gravitational field, the change in UCN energy between the top and bottom of a storage cell is about 31 neV, a significant fraction of the wall potential, 106.5 neV for  the liquid form of Fomblin used in \cite{mampe}.
Recent work has shown that surface capillary waves are a significant source of UCN loss due to inelastic upscattering beyond the wall potential\cite{lamgol}. The scattering will result in the change of the velocity spectrum of stored UCN. The analysis for the gravity correction that was performed in \cite{mampe} did not include these effects, for which the gravity correction possibly scales differently from the usual loss function.

Only the minimal published details of the experiment \cite{mampe} are available.   The lack of details and inconsistencies between the two publications \cite{mampe,nim} is astounding, and rather than adjust the model to exactly reproduce the results reported in those publications, a different approach was taken.  A model analysis was performed with and without the surface wave contribution, and the differences between the extrapolated lifetimes for the two cases were used to determine new correction terms to be added to those already included in \cite{mampe}.

In this calculation, the UCN energy will be given in cm, representing the change in energy with height of a UCN in the Earth's gravitational field.  The zero of energy will be take at the bottom of the storage cell.  Also, the neutron lifetime will be taken as infinity, so the correction to \cite{mampe} will be represented by any residual non-zero loss rate after the scaling reduction is applied to the numerical results.

In \cite{mampe}, the UCN storage cell was a box $H=30$ cm high, $W=40$ cm wide, with a variable length $L$ that was used to change the mean free path:
\begin{equation}
\lambda={2\over H^{-1}+W^{-1}+L^{-1}}.
\end{equation}
The UCN energy had a minimum value set by an Al window (53 cm potential) on the 6.4 cm diameter UCN guide, with a 20 cm rise to the storage cell.  The UCN enter the bottle at height about 10 cm from the bottom of the bottle, so the minimum UCN energy is taken as 43 cm.  There in no window on the separate guide that leads to the detector, so the minimum energy UCN detected will by 10 cm.
The results were insensitive to the assumed UCN energy spectrum, so a simple initial
\begin{equation}
n(E,0)=\sqrt{E} \ \ {\rm for}\ E>43\ {\rm cm}
\end{equation}
form for the initial density spectrum delivered to the bottle entrance was assumed.  However, in \cite{nim} it is stated that raising the storage cell by 20 cm resulted in a factor of three change in the number of stored UCN, suggesting a more significant high energy spectrum.  It is likely that the large change in the number of stored UCN with height is the result of an experimental artifact that is unknown. However, assuming a spectrum of form $n(E)=E^2$ did not significantly change the results of the calculations presented here.

An electrical analog model of the experimental system shown in Fig. 1 will help in the following discussion. The storage cell is represented by a capacitor that is charged (filled with UCN) by closing the switch located after $R_{in}$, and connects the capacitor to a current source.   A sliding door serves as the switch (valve), and in order for the scaling to work, the opening area is scaled (e.g., the valve is partially opened) with the storage volume so that the ratio of the impedance presented by the filling hole and guide ($R_{in}$) to the impedance representing the storage cell loss ($R_B$) is constant.  In the absence of gravity,
\begin{equation}
R_{in}(E)={4\over v(E) A_{in}}\ \ \ \ R_B(E)={4\over \mu_{tot}(E) v(E) A_B}
\end{equation}
where $v(E)=\sqrt{2gE}$, ($g=980 {\rm cm/s^2}$) is the UCN velocity for energy $E$, $A_{in}$ is the input valve opening area, $\langle \mu_{tot}(E)\rangle$ is the appropriately gravity average total UCN of energy E loss probability per reflection, and $A_B$ is the bottle area ($V_B$ is defined here as the bottle volume).  It is stated in \cite{mampe} that the fill hole opening area was varied to keep the loading of the source constant. Unfortunately, how this adjustment was performed is not given; one might assume that the sum fill hole and bottle impedances was kept constant; this requires that the fill hole be {\it reduced} for larger volumes. Although this keeps the spectrum constant within the guide, the effect on the neutron spectrum in the bottle itself is enormous. There is a further contradiction in \cite{nim} where it is stated that the filling time was scaled with $\lambda$. Changing the filling time cannot possibly work, so it is assumed here the filling area was varied.  This effect is referred to as ``loading the source" but it seems more profitable to think of the effect in terms of the relative impedances represented by the fill hole conductivity and the storage cell losses.  It should be noted, however, that there is a gravitational correction to $R_B$ that spoils the simple scaling. It would make more sense that the fill hole to the bottle area, $A_{in}/A_B$, and that is assumed in the calculations presented here.

In operation, the storage cell was filled for five or so net storage cell lifetimes, assuring equilibrium, after which the fill switch was opened (fill valve closed).  After a the storage period, the switch to the detector was closed (valve opened) and the UCN counted.  It is stated in \cite{mampe} that the guide lifetime was 15 s, but there is no statement of varying the opening area of the valve to the detector with storage cell volume.  It is stated in \cite{mampe} that the detector valve opening time was varied so that the number of collisions was constant over the experiment cycle.  Again, details of this time shift, possibly up to 30 seconds or so, are not described in any of the publications.  The numerical modelling was unstable regarding the form of the UCN detection, and shifts in apparent lifetime of up to 40 seconds were possible.  To get around this problem in the model presented here, the number of UCN left in the bottle at the end of a fixed storage period was determined, and as stated earlier, the change in number with and without the surface waves was used to determine a correction to the correction presented in \cite{mampe} so the results presented here are independent of these complicated and unknown details, and are expected to be a reasonably small percentage effect compared to what would be determined with a full model.

The average loss function is determined by the formalism presented in \cite{ucn,rp}.  First, the usual UCN loss due to absorption on the walls is given by
\begin{equation}
\mu=2f\left[{V\over E}\arcsin\left({E\over V}\right)^{1/2}-\left({V\over E}-1\right)^{1/2}\right]
\end{equation}
where $\mu$ is the loss probability due to nuclear absorption and molecular upscattering per surface collision, $f=8.3\times 10^{-6}$ is the ratio of the imaginary and real parts of the surface potential, and $V$ is the wall potential, approximately 104 cm for Fomblin, at 283 K \cite{lamgol}.
The effective volume of the bottle, in the presence of gravity, is given by
\begin{equation}
\gamma(E)=\int_0^{\min(E,H)} \sqrt{{E-h\over E}}\ \  A(h) dh
\end{equation}
where $A$ is the cross sectional area of the bottle at height $h$, constant and $A(h)=LW$ for a rectangular bottle, and we do not assume all neutrons in the storage cell can reach the roof (to allow for spectral evolution during storage).
The gravity averaged loss rate for UCN of energy E is then
\begin{equation}
\Gamma_w={1\over 4}{v(E)\over \gamma(E)}\int_0^{\min(E,H)} {E-h\over E}\ \  \mu(E-h)\ S(h)\ dh
\end{equation}
and
\begin{equation}
S(h)=LW(\delta(h)+\delta(h-H))+2W+2L
\end{equation}
where $\delta(x)$ is the Dirac delta function.

The surface wave upscattering/downscattering can be treated by a similar formalism.  From \cite{lamgol}, it can be seen that the probability to scatter a single UCN with a given initial energy to a final energy bin of width $\delta E_f$ is reasonably well described by
\begin{equation}
P(E_i\rightarrow E_f)\delta E_f=E_i(\alpha_1e^{-\beta_1 |E_i-E_f|)}+\alpha_2 e^{-\beta_2\sqrt|E_i-E_f|})\delta E_f
\end{equation}
where $\alpha_1=1\times 10^{-8}/$cm, $\beta_1=.065\ {\rm cm^{-1}} $, $\alpha_2=1\times 10^{-5}/$cm, and $\beta_2=2\ {\rm cm^{-1/2}}$ for Fomblin at 283 K.  Noting that the energy differences and bin widths are independent of $h$, the rate of change in spectrum can be calculated at $h=0$ using the same formalism as used with wall absorption to account for gravity, the gravity averaged rate of scattering from $E_i$ to $E_f\delta E_f$ is
\begin{equation}
\Gamma_{sw}(E_i\rightarrow E_f)\delta E_f=  {1\over 4}\ {P(E_i\rightarrow E_f)\delta E_f\over E_i}\ {v(E)\over \gamma(E)}\ \int_0^{\min(E,H)} {E_i-h\over E_i}\ S(h) \times  E_i \ dh.
\end{equation}
The spectral evolution from surface waves is then
\begin{eqnarray}
{dn(E,t)\over dt}&=&\int_0^V {\gamma(E_i)\over \gamma(E)}n(E_i,t)\Gamma_{sw}(E_i\rightarrow E)dE_i\\ &-&n(E,t)\left[\int_0^V \Gamma_{sw}(E\rightarrow E_f)+\int_V^\infty \Gamma_{sw}(E\rightarrow E_f)dE_f\right]\\ &=&{dn_{in}(E)\over dt}-n(E,t)\left[\Gamma_{sw,out}-\Gamma_{sw,loss}\right]\\
\end{eqnarray}
where the first term on the right hand side represents UCN scattered into final energy $E$ thus contributing to $n(E,t)$, the  second term UCN scattered out of energy $E$, and the third term represents UCN scattered from $E$ to energies higher than $V$ and are thus lost from the system. These equations conserve the total number of UCN in the system (total remaining stored plus those lost).

The equations were numerically integrated to determine the UCN spectrum and number density as a function of time.  The initial spectrum was taken as
\begin{equation}
n(E,0)=\sqrt{E} {\Gamma_{in}(E)\over \Gamma_{in}(E)+\Gamma_{sw,loss}(E)+\Gamma_w(E)}
\end{equation}
where
\begin{equation}
\Gamma_{in}(E)={1\over R_{in}(E) V_B}
\end{equation}
(the spectrum was not assumed to evolve due to the in,out scattering during filling). The spectrum was divided into bin 1 cm wide, and integrated in time over 1 s intervals:
\begin{equation}
n(E,t+\delta t)=n(E,t)+{dn_{in}(E,t)\over dt}\delta t -n(E,t)(\Gamma_w+\Gamma_{sw,out}+\Gamma_{sw,loss})\delta t
\end{equation}
and the total number of stored UCN that can be detected as a function of time is
\begin{equation}
N(t)=\int_{10}^V \gamma(E) n(E,t) dE.
\end{equation}

Results for the spectral evolution for initially monochromatic UCN of 70 cm energy, with and without gravity, are shown in Fig. 2.  Results of scaled extrapolation of the neutron lifetime however indicated an effect of less than 0.5 sec on the change in the neutron lifetime.  Figure 3 shows typical data that was used in this extrapolation. Numerous tests for effects of initial spectrum, bottle emptying procedure, and source loading, were performed.  It might be surprising that such a large effect on the UCN spectrum does not lead to a large correction, but insofar as the gravity correction follows the scaling law as used in \cite{mampe}, the effect should not contribute.  Indeed, shown in Fig. 4 are the results for storage at time/mfp scaled calculations. It can be seen that the effects of gravity are negligible, in that there is a less than 1\% difference in the spectra, and the majority of the spectrum remains with $\pm 10$ cm of the initial energy of 70 cm.  Reasonable agreement with the corrections given in \cite{mampe} was obtained. In addition, the effects of the surface wave was determined by adjusting $f$ for the cases with and without surface wave to give approximately the same loss rate.

With that said, it was not possible to fully model the experiment described in \cite{mampe} with the available information.  This study would be better done in regard to the  experiment described in \cite{aruz}, which appears as more complicated, so the simple electrical models used here are not applicable.  However, there are enough experimental details that exist in Ph.D. dissertations and published papers to allow construction of a reasonable model of the apparatus described in \cite{aruz}.

\begin{figure}
\begin{center}
\includegraphics[
width=4in ] {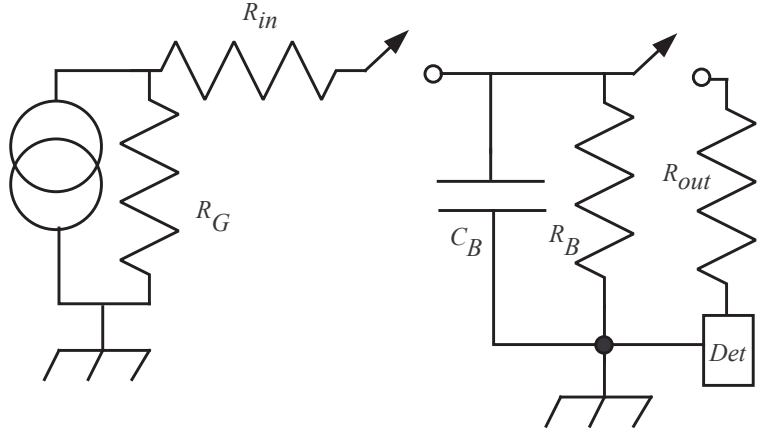} \caption{Electrical circuit analog of the neutron storage experiment.}
\end{center}
\end{figure}

\begin{figure}
\begin{center}
\includegraphics[
width=4in ] {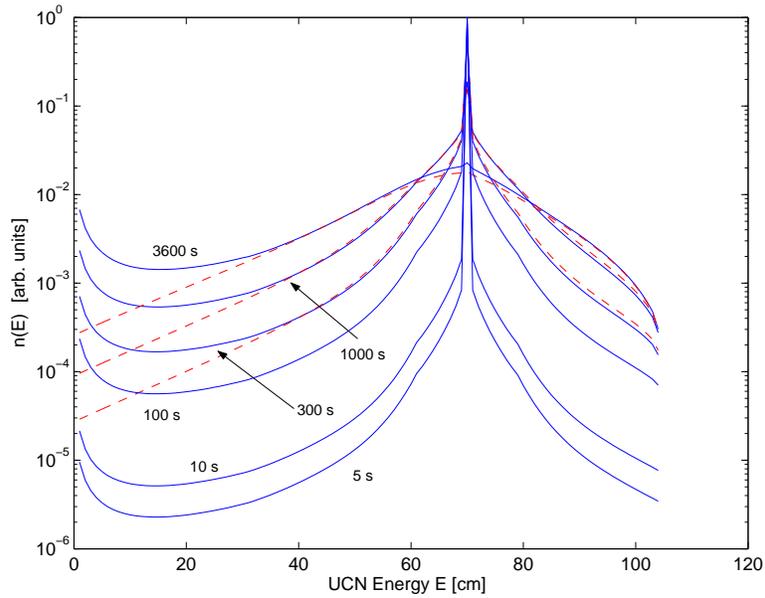} \caption{Evolution of an initially monochromatic UCN spectrum (70 cm) due to surface wave scattering, $L=85$ cm.  Loss due to upscattering and wall absorption are included, but the beta-decay rate is set to zero (as it is for all calculations in this note).}
\end{center}
\end{figure}

\begin{figure}
\begin{center}
\includegraphics[
width=4in ] {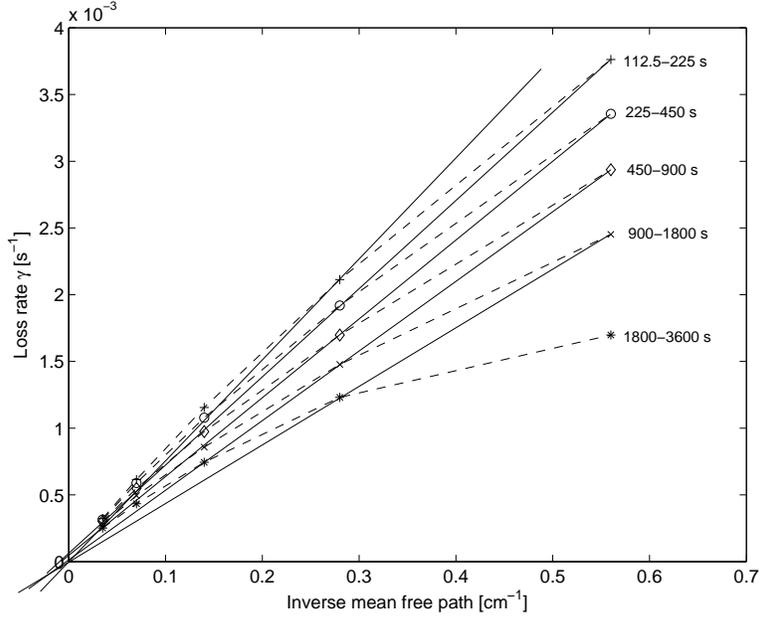} \caption{Lifetime/mean free path scaled numerical results.  Extrapolation to infinite mean free path determines a gravitational correction factor. Effects of finite numerical precision were made negligible by considering cases with and without gravity, and with and without surface waves.}
\end{center}
\end{figure}

\begin{figure}
\begin{center}
\includegraphics[
width=4in ] {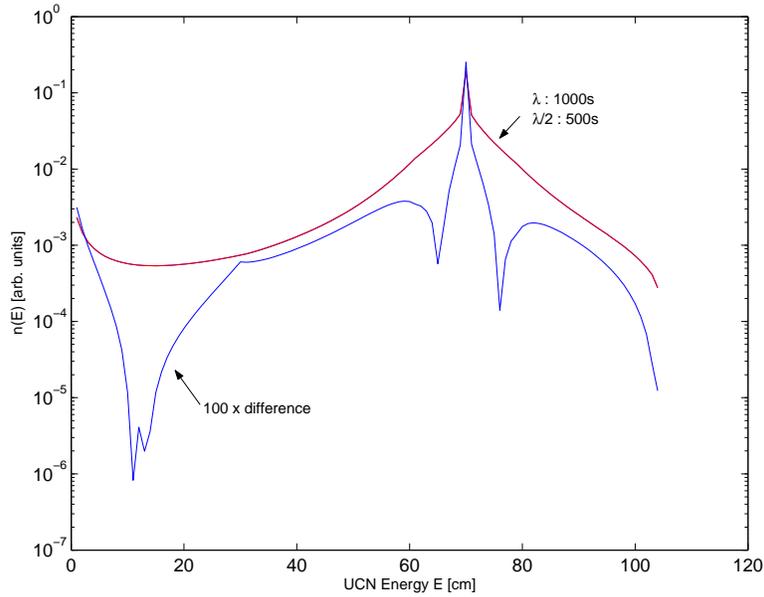} \caption{Spectrum for two cases: $L=85.71$ cm, $t=1000$ s compared with $L=12.24$ cm (one-half mfp), $t=500$ s. The differences are small, and can be used to explain the fact that the surface wave scattering does not affect the scaling extrapolated neutron lifetime.}
\end{center}
\end{figure}

\end{document}